\documentclass[aps,prl,twocolumn]{revtex4}
\usepackage{graphicx}
\usepackage{dcolumn}
\usepackage{bm}
\usepackage{amssymb}

\begin{document}
%%%%%%%%%%%%%%%%%%%%%%%%%%%%%%%%%%%%%%%%%%%%%%%%%%%%%%%%%%%%%%%%%%%%%%%%%%%%%%%%
%%%%%%%%%%%%%%%%%%%%%%%%%%%%%%%%%%%%%%%%%%%%%%%%%%%%%%%%%%%%%%%%%%%%%%%%%%%%%%%%

\title{Few-Photon All-Optical $\pi$ Phase Modulation Based on a Double-$\Lambda$ System}

\author{Yen-Chun Chen,$^{1,\dag}$ Hao-Chung Chen,$^1$ Hsiang-Yu Lo,$^{1,\ddag}$ Bing-Ru Tsai,$^1$ Ite A. Yu,$^2$ Ying-Cheng Chen,$^3$ and Yong-Fan Chen,$^1$}

\email{yfchen@mail.ncku.edu.tw}

\altaffiliation{$^\dag$Present address: Department of Electrophysics, National Chiao Tung University, Hsinchu 30013, Taiwan}
\altaffiliation{$^\ddag$Present address: Institute for Quantum Electronics, ETH Z\"{u}rich, 8093 Z\"{u}rich, Switzerland}

\affiliation{Department of Physics, National Cheng Kung University, Tainan 70101, Taiwan \\
$^2$Department of Physics and Frontier Research Center on Fundamental and Applied Sciences of Matters, National Tsing Hua
University, Hsinchu 30013, Taiwan \\
$^3$Institute of Atomic and Molecular Sciences, Academia Sinica, Taipei 10617, Taiwan}

%\date{\today}

%%%%%%%%%%%%%%%%%%%%%%%%%%%%%%%%%%%%%%%%%%%%%%%%%%%%%%%%%%%%%%%%%%%%%%%%%%%%%%%%
%%%%%%%%%%%%%%%%%%%%%%%%%%%%%%%%%%%%%%%%%%%%%%%%%%%%%%%%%%%%%%%%%%%%%%%%%%%%%%%%

\begin{abstract}

We propose an efficient all-optical phase modulation based on a double-$\Lambda$ system and demonstrate a $\pi$ phase shift of a few-photon
pulse induced by another few-photon pulse in cold rubidium atoms with this scheme. By changing the phases of the applied laser fields, we can
control the property of the double-$\Lambda$ medium. This phase-dependent mechanism makes the double-$\Lambda$ system different form the
conventional cross-Kerr-based system which only depends on the applied laser intensities. The proposed scheme provides a new route to generate
strong nonlinear interactions between photons, and may have potential for applications in quantum information technologies.

\end{abstract}

%%%%%%%%%%%%%%%%%%%%%%%%%%%%%%%%%%%%%%%%%%%%%%%%%%%%%%%%%%%%%%%%%%%%%%%%%%%%%%%%
%%%%%%%%%%%%%%%%%%%%%%%%%%%%%%%%%%%%%%%%%%%%%%%%%%%%%%%%%%%%%%%%%%%%%%%%%%%%%%%%

%03.67.-a Quantum information
%32.80.Qk Coherent control of atomic interactions with photons
%42.25.Hz Interference
%42.50.Gy Effects of atomic coherence on propagation, absorption, and amplification of light; electromagnetically induced transparency and absorption

%%%%%%%%%%%%%%%%%%%%%%%%%%%%%%%%%%%%%%%%%%%%%%%%%%%%%%%%%%%%%%%%%%%%%%%%%%%%%%%%
%%%%%%%%%%%%%%%%%%%%%%%%%%%%%%%%%%%%%%%%%%%%%%%%%%%%%%%%%%%%%%%%%%%%%%%%%%%%%%%%

\maketitle

%%%%%%%%%%%%%%%%%%%%%%%%%%%%%%%%%%%%%%%%%%%%%%%%%%%%%%%%%%%%%%%%%%%%%%%%%%%%%%%%
%%%%%%%%%%%%%%%%%%%%%%%%%%%%%%%%%%%%%%%%%%%%%%%%%%%%%%%%%%%%%%%%%%%%%%%%%%%%%%%%
\newcommand{\FigOne}{
    \begin{figure}[t] %Fig.1
    \includegraphics[width=9.00cm]{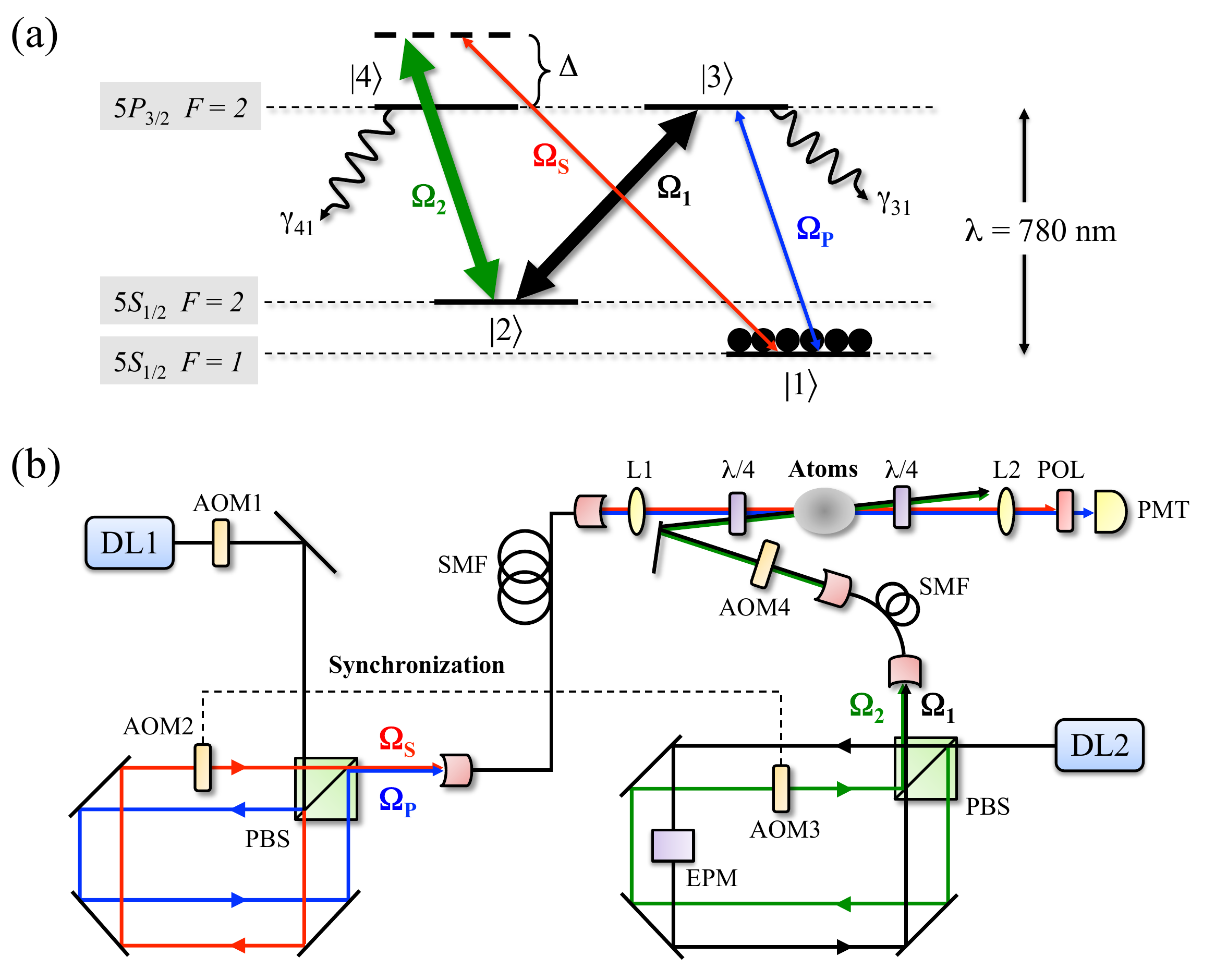}
    \caption{(color online).
Energy level scheme and experimental apparatus. (a) Energy levels of $^{87}{\rm Rb}$ $D_2$-line transition for the double-$\Lambda$ EIT
experiment. Signal detuning, $\Delta$, is defined as $\omega_s - \omega_{24}$, where $\omega_s$ and $\omega_{24}$ are the frequencies of the
signal field and the $|2\rangle \leftrightarrow |4\rangle$ transition, respectively. (b) Schematic diagram of the experimental setup. DL, diode
laser; PBS, polarizing-beam splitter; AOM, acousto-optic modulator; $\lambda$/4, quarter-wave plate; POL, polarizer; SMF, single-mode fiber; L,
Lens; EPM, electro-optic phase modulator; PMT, photomultiplier tube.}
    \label{fig:setup}
    \end{figure}
}
%%%%%%%%%%%%%%%%%%%%%%%%%%%%%%%%%%%%%%%%%%%%%%%%%%%%%%%%%%%%%%%%%%%%%%%%%%%%%%%%
\newcommand{\FigTwo}{
    \begin{figure}[t] %Fig.2
    \includegraphics[width=9.00cm]{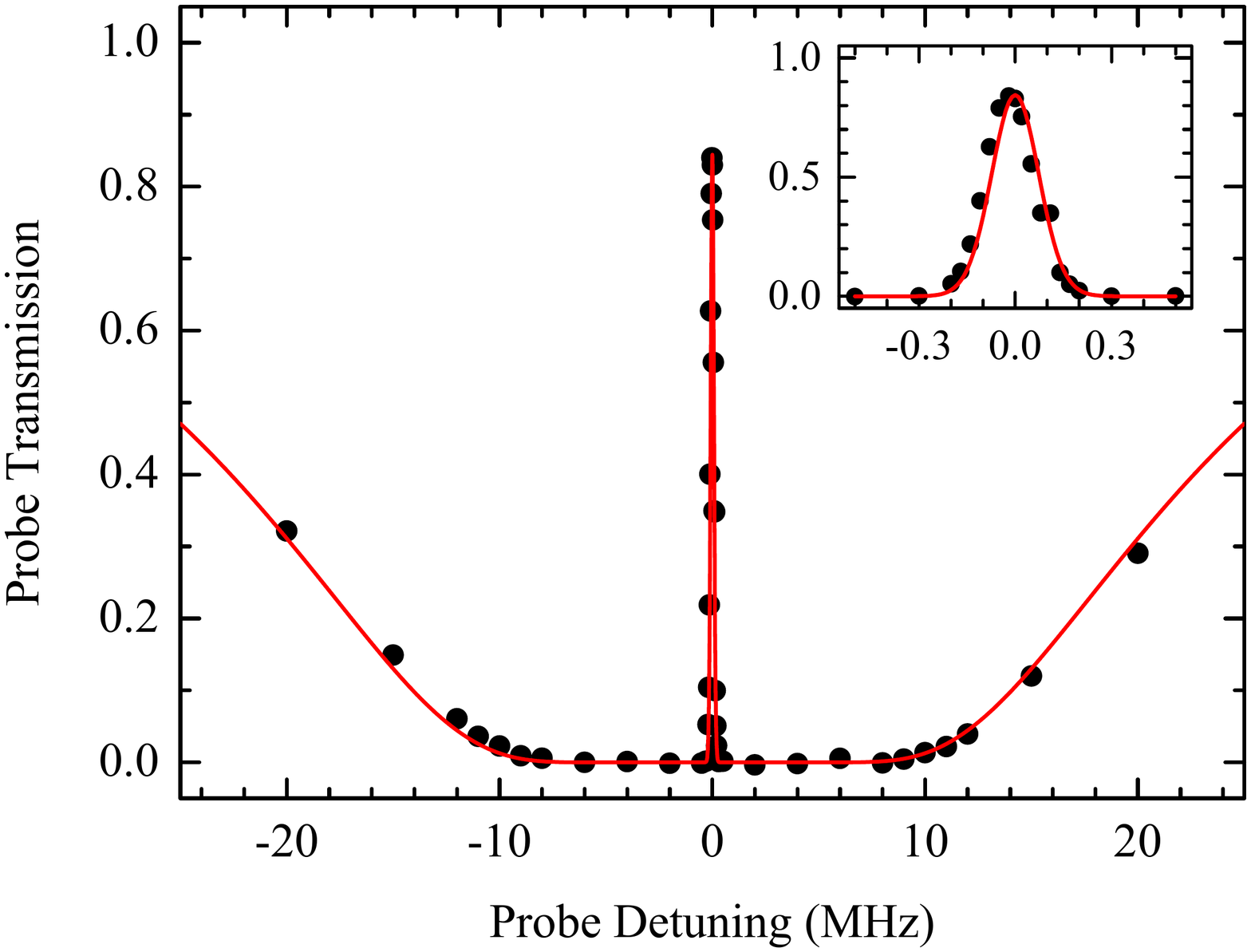}
    \caption{(color online).
Observed EIT transmission versus probe field detuning. The black circles and red line represent the measurement data and
theoretical curve, respectively. The inset shows the EIT transmission window. Probe detuning is defined as $\omega_p -
\omega_{13}$, where $\omega_p$ and $\omega_{13}$ are the frequencies of the probe field and the $|1\rangle \leftrightarrow
|3\rangle$ transition, respectively. The parameters for the theoretical curve are $\alpha_{p}=34$, $|\Omega_1|=0.5\Gamma$,
$\gamma_{21}=0.001\Gamma$, and $\gamma_{31}=\gamma_{41}=1.25\Gamma$.}
    \label{fig:forward}
    \end{figure}
}
%%%%%%%%%%%%%%%%%%%%%%%%%%%%%%%%%%%%%%%%%%%%%%%%%%%%%%%%%%%%%%%%%%%%%%%%%%%%%%%%
\newcommand{\FigThree}{
    \begin{figure}[t] %Fig.3
    \includegraphics[width=9.00cm]{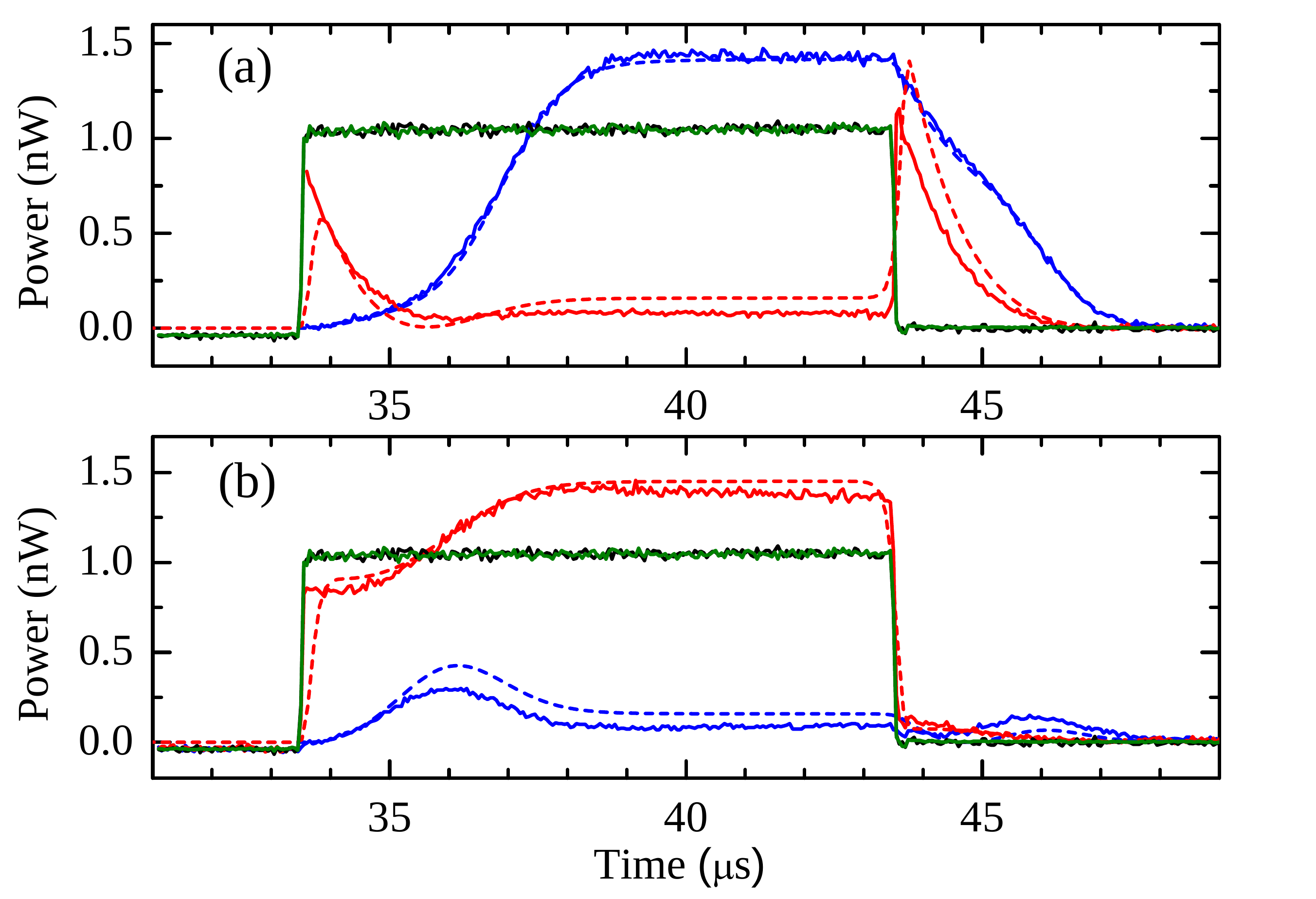}
    \caption{(color online).
Double-$\Lambda$ experiment in the pulsed regime. The solid and dashed lines represent the experimental data and theoretical
curves, respectively. The black (green) lines are the input probe (signal) pulses; the blue (red) are the transmitted probe
(signal) pulses. The parameters for the theoretical curves (dashed lines) are $\alpha_{p}=46$, $|\Omega_1| = |\Omega_2| =
0.7\Gamma$, $\Delta = 13\Gamma$, $\gamma_{21}=0.001\Gamma$, $\gamma_{31}=\gamma_{41}=1.25\Gamma$. (a) the relative phase
$\phi_{r}$ = 1.5 rad. (b) $\phi_{r}$ = 4.5 rad. The peak powers of the probe and signal pulses were 1 nW, corresponding to 40,000
photons per pulse.}
    \label{fig:beat}
    \end{figure}
}
%%%%%%%%%%%%%%%%%%%%%%%%%%%%%%%%%%%%%%%%%%%%%%%%%%%%%%%%%%%%%%%%%%%%%%%%%%%%%%%%
\newcommand{\FigFour}{
    \begin{figure}[t] %Fig.4
    \includegraphics[width=9.2cm]{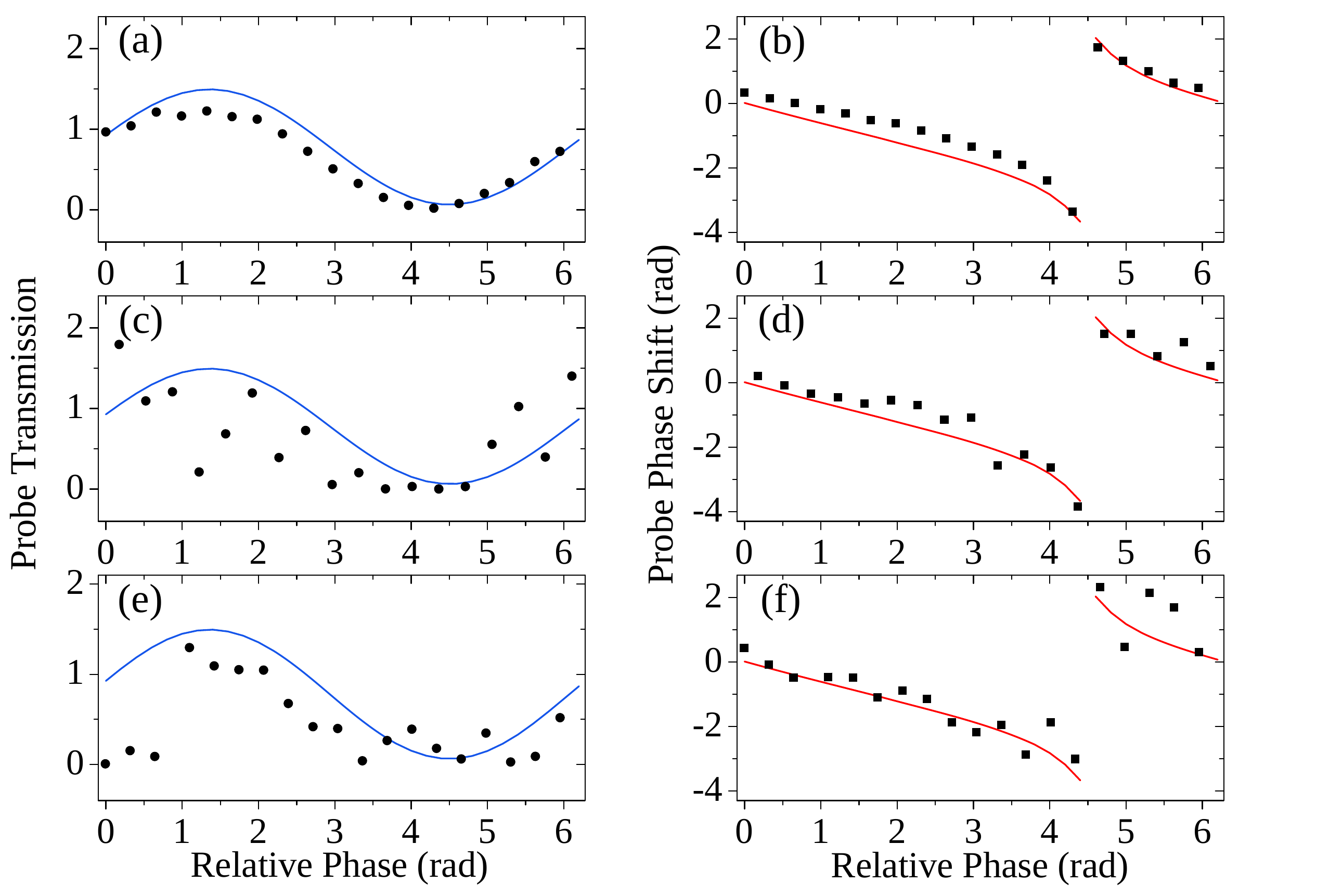}\label{fig:fig4}
    \caption{(color online).
Few-photon all-optical phase modulation. Dependence of the transmission and the phase shift of the probe pulse on the relative
phase $\phi_{r}$. The numbers of both the probe and the signal photons are 40,000 in (a) and (b); 40 in (c) and (d); 16 in (e) and
(f). Circles (squares) represent the transmission (phase shift) of the probe pulse. The blue and red lines are the theoretical
curves of the probe transmission and phase shift, respectively. The parameters for the theoretical curves are $\alpha_{p}=41$,
$|\Omega_1| = |\Omega_2| = 0.7\Gamma$, $\Delta = 13\Gamma$, $\gamma_{21}=0.001\Gamma$, $\gamma_{31}=\gamma_{41}=1.25\Gamma$.}
    \label{fig:backward}
    \end{figure}
}
%%%%%%%%%%%%%%%%%%%%%%%%%%%%%%%%%%%%%%%%%%%%%%%%%%%%%%%%%%%%%%%%%%%%%%%%%%%%%%%%
%%%%%%%%%%%%%%%%%%%%%%%%%%%%%%%%%%%%%%%%%%%%%%%%%%%%%%%%%%%%%%%%%%%%%%%%%%%%%%%%

%\section{Abstract and Introduction}

The realization of all-optical $\pi$ phase modulation, ultimately at the single-photon level, is a challenging task in quantum information
science~\cite{TurchetteCavityXPM, FushmanCavityXPM, MatsudaFiberXPM}. To enhance strong optical nonlinearities at low light levels,
electromagnetically induced transparency (EIT) is one of the most promising technologies~\cite{HarrisEIT, LukinEIT, FleischhauerEIT}. An N-type
all-optical control of light based on EIT has attracted considerable attention over the past decades~\cite{SchmidtXPM, HarrisPS, HarrisSLNO,
HauSlowLt, LukinCP, KangXPM, BrajePS, ChenPS, LoXPM}. Recently, Bajcsy~\textit{et al.} have realized the N-type all-optical switching in an
atom-filled hollow fiber with light pulses containing a few hundred photons~\cite{BajcsyFiberPS}. Furthermore, the N-type all-optical phase
modulation (APM) at the level of a few hundred photons has also been demonstrated in cold atoms by Lo~\textit{et al.}~\cite{LoFXPM}. However,
due to the interaction of light pulses with unequal group velocities, the efficiency of the N-type APM has an upper limit of order 0.1 radians
at the single-photon level, as investigated by Harris~\textit{et al.}~\cite{HarrisSLNO}. To overcome this upper limit, there have been many
theoretical proposals and experimental studies on this subject in recent years~\cite{LukinDSLXPM, ChenLSXPM, ShiauDSXPM, YuSLXPM}. These methods
often require a beam tightly focused to a spot size equal to the area of the atomic absorption cross section as well as a medium with a large
optical depth, which remain technically challenging.

%Although there are several proposed methods to enhance optical nonlinear effects, the interaction between two light fields is normally too small
%for practical applications.

Here we propose a novel scheme of APM based on a phase-dependent double-$\Lambda$ EIT system and demonstrate a large phase shift of a few-photon
pulse induced by another few-photon pulse in cold $^{87}{\rm Rb}$ atoms. A nonlinear phase shift of order $\pi$ induced by a light pulse
containing 16 photons was observed with this scheme, to our knowledge, which is currently the best result in EIT-based APM schemes.
Additionally, a gain behavior resulting from coherent photon transfer due to the competition between two four-wave mixing (FWM) processes was
observed in this double-$\Lambda$ system.

%%%%%%%%%%%%%%%%%%%%%%%%%%%%%%%%%%%%%%%%%%%%%%%%%%%%%%%%%%%%%%%%%%%%%%%%%%%%%%%%
%%%%%%%%%%%%%%%%%%%%%%%%%%%%%%%%%%%%%%%%%%%%%%%%%%%%%%%%%%%%%%%%%%%%%%%%%%%%%%%%

%\section{Experimental Details}

In the present study, we investigated the double-$\Lambda$-based APM in a laser-cooled $^{87}{\rm Rb}$ atomic system, as depicted in Fig.~1(a).
Dense cold atomic gas with a large optical depth of approximately 40 was produced in a dark spontaneous-force optical trap
(SPOT)~\cite{KetterleSPOT}. The dark SPOT was implemented using a typical magneto-optical trap with a bright capturing region, and two
perpendicular repumping beams (diameter 2.5 cm, power 0.4 mW) with 5 mm diameter dark areas that drove the $|1\rangle \leftrightarrow |3\rangle$
transition resonantly to form a dark region in the center of the trap. The temperature of the cold $^{87}{\rm Rb}$ atoms produced in the dark
SPOT was around 300 $\mu$K.

A strong coupling field ($\Omega_1$ denotes its Rabi frequency) drove the $|2\rangle \leftrightarrow |3\rangle$ transition to create a
transparent medium for a weak probe pulse ($\Omega_p$, $|1\rangle \leftrightarrow |3\rangle$) through quantum interference. The coupling and
probe fields formed the first $\Lambda$-type EIT system. The second $\Lambda$-type EIT system was created by a strong control field ($\Omega_2$,
$|2\rangle \leftrightarrow |4\rangle$) and a weak signal pulse ($\Omega_s$, $|1\rangle \leftrightarrow |4\rangle$). In the experiment, the
coupling and probe fields were right circularly polarized ($\sigma+$) while the control and signal fields were left circularly polarized
($\sigma-$). The four laser fields drove the $D_2$-line transition of the $^{87}{\rm Rb}$ atoms to form the double-$\Lambda$ EIT system, as
shown in Fig.~1(a).

A schematic diagram of the experimental setup is shown in Fig.~1(b). The probe and signal fields were produced using a single
diode laser (DL1); the coupling and control fields were produced using another diode laser (DL2). DL2 was directly injection
locked using an external cavity diode laser (ECDL, TOPTICA DL 100) with a laser linewidth of around 1 MHz. One beam from the ECDL
was sent through a 6.8-GHz electro-optic modulator (EOM, New Focus 4851). DL1 was injection locked by an intermediate laser seeded
with the high-frequency sideband of the EOM output. The above arrangement is capable of completely eliminating the influence of
the carrier of the EOM output on DL1. The probe beam was overlapped with the signal beam on a polarization beam splitter (PBS),
and then sent to a single-mode fiber (SMF) to obtain the optimal spatial mode-matching. The $e^{-2}$ diameters of the probe
(signal) and coupling (control) beams were 0.2 mm and 3 mm, respectively. These two beams propagated at an angle of $0.9^\circ$.
All of the laser fields were switched on and off via acousto-optic modulators (AOMs). We utilized AOM1 to control the widths of
the probe and signal pulses. The coupling and control fields were switched on and off via AOM4, as shown in Fig.~1(b). The
experimental data were detected by a photomultiplier tube module (PMT, Hamamatsu H6780-20 and C9663) with a conversion gain of $9
\times 10^{7}$ V/W, and then recorded using an oscilloscope (Agilent MSO6034A) throughout the experiment.

When conducting the phase-dependent double-$\Lambda$ experiment, an electro-optic phase modulator (EPM, Thorlabs EO-PM-NR-C1) was applied to
vary the phase of the coupling field ($\Omega_1$). Furthermore, to stabilize the relative phase of the four laser fields, two main setups were
utilized in this experiment. (i) The optical paths of the probe and signal (coupling and control) fields were arranged in the configuration of a
Sagnac-like interferometer to reduce the path fluctuations between these two beams, as shown in Fig.~1(b). (ii) AOM2 and AOM3 were driven by the
same RF generator through an RF power splitter (Mini-Circuits ZMSC-2-1+). However, due to temperature fluctuations, there was a long-term phase
drift of approximately 1 radian per hour in the current experiment.

%%%%%%%%%%%%%%%%%%%%%%%%%%%%%%%%%%%%%%%%%%%%%%%%%%%%%%%%%%%%%%%%%%%%%%%%%%%%%%%%
%%%%%%%%%%%%%%%%%%%%%%%%%%%%%%%%%%%%%%%%%%%%%%%%%%%%%%%%%%%%%%%%%%%%%%%%%%%%%%%%

%\section{Theoretical Model}

To theoretically analyze the behavior of the probe and signal pulses propagating in the double-$\Lambda$ EIT medium, we used the
Maxwell-Schrodinger equations below:
\begin{eqnarray}
\frac{\partial\Omega_{p}}{\partial z} + \frac{1}{c}\frac{\partial\Omega_{p}}{\partial t} &= i \frac{\alpha_{p}\gamma_{31}}{2L}
\rho_{31},
\end{eqnarray}
\begin{eqnarray}
\frac{\partial\Omega_{s}}{\partial z} + \frac{1}{c}\frac{\partial\Omega_{s}}{\partial t} &= i \frac{\alpha_{s}\gamma_{41}}{2L}
\rho_{41},
\end{eqnarray}
where $\Omega_p = |\Omega_p| e^{i\phi_{p}}$ and $\Omega_s = |\Omega_s| e^{i\phi_{s}}$ are the Rabi frequencies of the probe and signal pulses,
respectively. $\phi_{p}$ ($\phi_{s}$) describes the phase information carried by the probe (signal) pulse. $\rho_{31}$ ($\rho_{41}$) is the
slowly-varying amplitude of the optical coherence of the probe (signal) transition. $\alpha_{p} = n \sigma_{13}L$ ($\alpha_{s} = n\sigma_{14}L$)
represents the optical depth of the probe (signal) transition, where $n$ is the number density of the atoms, $\sigma_{13}$($\sigma_{14}$) is the
atomic absorption cross section of the probe (signal) transition, and $L$ is the optical path length of the medium. $\gamma_{31}$ and
$\gamma_{41}$ are the total coherence decay rates from the $|3\rangle$ and $|4\rangle$ excited states, respectively. We note that the optical
depths of the probe and signal transitions in this experiment are the same ($\alpha_{p}=\alpha_{s}$) because $\sigma_{13}$ is equal to
$\sigma_{14}$ by considering three degenerate Zeeman sublevels, as shown in Fig.~1(a).

%%%%%%%%%%%%
\FigOne
%%%%%%%%%%%%

In the case where the probe and signal fields are very weak (i.e., $\rho_{11} \simeq 1$), the optical Bloch equations of the slowly-varying
amplitudes of the density-matrix elements are given by:
\begin{eqnarray}
\frac{d}{dt}\rho_{41} &= \frac{i}{2}\Omega_{s} + \frac{i}{2}\Omega_{2}\rho_{21} + \left(i\Delta -
\frac{\gamma_{41}}{2}\right)\rho_{41},
\end{eqnarray}
\begin{eqnarray}
\frac{d}{dt}\rho_{31} &= \frac{i}{2}\Omega_{p} + \frac{i}{2}\Omega_{1}\rho_{21} - \frac{\gamma_{31}}{2}\rho_{31},
\end{eqnarray}
\begin{eqnarray}
\frac{d}{dt}\rho_{21} &= \frac{i}{2}\Omega^{\ast}_{1}\rho_{31} + \frac{i}{2}\Omega^{\ast}_{2}\rho_{41} -
\frac{\gamma_{21}}{2}\rho_{21},
\end{eqnarray}
where $\Omega_1 = |\Omega_1| e^{i\phi_{1}}$ and $\Omega_2 = |\Omega_2| e^{i\phi_{2}}$ are the Rabi frequencies of the coupling and control
transitions, respectively. $\phi_{1}$ ($\phi_{2}$) describes the phase information carried by the coupling (control) field. $\Delta$ denotes the
detuning of the signal transition [see Fig.~1(a)]. $\gamma_{21}$ represents the dephasing rate of the $|1\rangle$ and $|2\rangle$ ground states.
Each parameter in the theoretical model was individually determined from additional experiments as follows: $|\Omega_1|$ was determined from the
separation of the two absorption peaks in the EIT spectrum. $|\Omega_2|$ was determined from the EIT-based photon switching~\cite{ChenPS}.
$\alpha_{p}$ was derived from the delay time of the slow light pulse~\cite{HauSlowLt}. $\gamma_{21}$ was $0.0010(2)\Gamma$, as estimated by the
degree of EIT transparency. $\Gamma = 2\pi \times 6$ MHz is the spontaneous decay rate of the excited states. $\gamma_{31}$ and $\gamma_{41}$
were both $1.25(2)\Gamma$, contributed mostly by the spontaneous decay rate and laser linewidth, as obtained from the spectral width of the
one-photon absorption. Under the conditions of $\gamma_{21} = 0$ and $\gamma_{31}$ = $\gamma_{41}$, the steady-state solutions of Eqs.~(1)-(5)
for the probe and signal fields are:

\[
\Omega_p(\alpha_{p}) = \frac{|\Omega_1|}{|\Omega|^{2}}
{\left[|\Omega_1||\Omega_p(0)|e^{i\phi_{p}}+|\Omega_2||\Omega_s(0)|e^{i\left(\phi_{p}-\phi_{r}\right)}\right]}+
\]
\begin{equation}
\frac{|\Omega_2|}{|\Omega|^{2}}
{\left[|\Omega_2||\Omega_p(0)|e^{i\phi_{p}}-|\Omega_1||\Omega_s(0)|e^{i\left(\phi_{p}-\phi_{r}\right)}\right]}e^{-i\frac{\alpha_{p}}{2\xi}},
\end{equation}
\\
\[
\Omega_s(\alpha_{s}) = \frac{|\Omega_2|}{|\Omega|^{2}}
{\left[|\Omega_2||\Omega_s(0)|e^{i\phi_{s}}+|\Omega_1||\Omega_p(0)|e^{i\left(\phi_{s}+\phi_{r}\right)}\right]}+
\]
\begin{equation}
\frac{|\Omega_1|}{|\Omega|^{2}}
{\left[|\Omega_1||\Omega_s(0)|e^{i\phi_{s}}-|\Omega_2||\Omega_p(0)|e^{i\left(\phi_{s}+\phi_{r}\right)}\right]}e^{-i\frac{\alpha_{s}}{2\xi}},
\end{equation}
where $|\Omega|^{2}=|\Omega_1|^{2}+|\Omega_2|^{2}$, $\xi=i+2\frac{|\Omega_1|^{2}\Delta}{|\Omega|^{2}\gamma_{31}}$, and we define the relative
phase of the four laser fields, $\phi_{r}$, as ($\phi_{p}$-$\phi_{1}$)-($\phi_{s}$-$\phi_{2}$). In the case where $\Delta=0$,
$|\Omega_p(0)|=|\Omega_s(0)|$, and $|\Omega_1|=|\Omega_2|$, according to Eqs.~(6) and (7), $\Omega_p(\alpha_{p})=|\Omega_p(0)|e^{i\phi_{p}}$ and
$\Omega_s(\alpha_{s})=|\Omega_s(0)|e^{i\phi_{s}}$ when $\phi_{r}=0$, which means the double-$\Lambda$ medium becomes completely transparent for
both the probe and the signal fields. On the other hand, when $\phi_{r}=\pi$, the medium becomes opaque and has maximum attenuation for both the
probe and the signal fields. This phase-dependent double-$\Lambda$ system with $\Delta = 0$ can be applied in low-light-level all-optical
switching, as previously described~\cite{ZhuDLambdaPS1}. Here we focus on demonstrating all-optical phase modulation based on a non-resonant
double-$\Lambda$ system with $\Delta \neq 0$. Of note, the influence of the relative phase of the applied laser fields on the property of the
double-$\Lambda$ medium has been theoretically discussed in Ref.~\cite{KorsunskyDLambdaT}. The matched propagation of a pair of ultraslow light
pulses in the double-$\Lambda$ system has also been studied in Ref.~\cite{DengDLambdaT}.

%%%%%%%%%%%%
\FigTwo
%%%%%%%%%%%%

%%%%%%%%%%%%%%%%%%%%%%%%%%%%%%%%%%%%%%%%%%%%%%%%%%%%%%%%%%%%%%%%%%%%%%%%%%%%%%%%
%%%%%%%%%%%%%%%%%%%%%%%%%%%%%%%%%%%%%%%%%%%%%%%%%%%%%%%%%%%%%%%%%%%%%%%%%%%%%%%%

% Experimental result & discussion - EIT Spectrum

We first measured the transmission of a probe pulse propagating through a three-level $\Lambda$-type EIT medium. After all of the
lasers and magnetic fields of the dark SPOT were turned off and the coupling field ($\Omega_1$) was switched on for 100 $\mu$s,
the 50-$\mu$s probe square pulse was switched on to perform the measurement. The experiment was conducted at a repetition rate of
100 Hz. The input power of the probe pulse was set to 1 nW in the EIT experiment. The Rabi frequency of the coupling transition,
$|\Omega_1|$, was 0.5$\Gamma$, corresponding to the coupling laser power of 258 $\mu$W. Figure 2 shows the probe transmission as a
function of probe detuning. The inset shows the EIT transmission window. The measurement data (circles) are in good agreement with
the theoretical curve (red line). The theoretical curve was plotted using the EIT theoretical expression in
Ref.~\cite{LoBeatNote}.

%%%%%%%%%%%%
\FigThree
%%%%%%%%%%%%

%%%%%%%%%%%%%%%%%%%%%%%%%%%%%%%%%%%%%%%%%%%%%%%%%%%%%%%%%%%%%%%%%%%%%%%%%%%%%%%%
%%%%%%%%%%%%%%%%%%%%%%%%%%%%%%%%%%%%%%%%%%%%%%%%%%%%%%%%%%%%%%%%%%%%%%%%%%%%%%%%

% Experimental result & discussion - Double-\Lambda XPM in the pulsed regime

Next, a non-resonant double-$\Lambda$ experiment in the pulsed regime was performed. Figure 3 shows typical data from the double-$\Lambda$
experiment, where $\alpha_{p}=46$, $\Delta$ = 13$\Gamma$, $|\Omega_1| = |\Omega_2| = 0.7\Gamma$, and the input powers of both the probe and the
signal pulses were set to 1 nW, corresponding to $|\Omega_p(0)|=|\Omega_s(0)|= 0.016\Gamma$ (i.e., $|\Omega_{p(s)}(0)| \ll |\Omega_{1(2)}|$).
Here the widths of both the probe and signal pulses were set to 10 $\mu$s. We utilized the EPM to vary the relative phase $\phi_{r}$ in the
double-$\Lambda$ experiment [see Fig.~1(b)]. The relative phase $\phi_{r}$ was set to 1.5 radians in Fig.~3(a) and 4.5 radians in Fig.~3(b). The
solid and dashed lines represent the experimental data and theoretical curves, respectively. The theoretical curves were plotted by numerically
solving Eqs.~(1)-(5). The black (green) lines are the input probe (signal) pulses, and the blue (red) lines are the transmitted probe (signal)
pulses. The group-velocity mismatch of the transmitted probe and signal pulses in Fig.~3 is due to $\Delta \neq 0$. The experimental data also
show that the power of the transmitted light exceeds its input power in the non-resonant double-$\Lambda$ system. This gain behavior of the
double-$\Lambda$ medium resulted from coherent photon transfer between two N-type FWM processes ($|1\rangle \rightarrow |3\rangle \rightarrow
|2\rangle \rightarrow |4\rangle \rightarrow |1\rangle$ and $|1\rangle \rightarrow |4\rangle \rightarrow |2\rangle \rightarrow |3\rangle
\rightarrow |1\rangle$). It is worth noting that recently the techniques of FWM-based coherent photon conversion in photonic crystal fibres have
been proposed to implement an efficient quantum computing~\cite{LangfordCPC}. Furthermore, McGuinness~\textit{et al.} have demonstrated that
FWM-based quantum frequency conversion can preserve the number statics of single-photon states~\cite{McGuinnessQFC,RaymerCPC}.

%The experimental data are in agreement with the theoretical curves.

%%%%%%%%%%%%%%%%%%%%%%%%%%%%%%%%%%%%%%%%%%%%%%%%%%%%%%%%%%%%%%%%%%%%%%%%%%%%%%%%
%%%%%%%%%%%%%%%%%%%%%%%%%%%%%%%%%%%%%%%%%%%%%%%%%%%%%%%%%%%%%%%%%%%%%%%%%%%%%%%%

% Experimental result & discussion - Few-photon all-optical \pi phase modulation

To measure the phase shift of the weak probe pulse induced by the weak signal pulse, we utilized a beat-note interferometer for low-light-level
phase measurement in the pulsed regime. The phase shift of the probe pulse was measured by directly comparing the reference and probe beat
notes. The probe transmission was simultaneously obtained from the amplitude of the probe beat notes. The experimental setup and details of the
beat-note interferometer can be found in Ref.~\cite{LoBeatNote}.

Figure 4 shows the experimental data of the double-$\Lambda$-based APM. The experimental parameters are the same as those in Fig.~3 except for
the optical depth ($\alpha_{p}=41$). We first performed the double-$\Lambda$ experiment where the input powers of both the probe and the signal
pulses were set to 1 nW, corresponding to 40,000 photons per pulse. Figure 4(a) and 4(b) show the experiment results of the dependence of the
probe transmission and phase shift on the relative phase $\phi_{r}$, respectively. Next, we performed the double-$\Lambda$ experiment at the
few-photon level. The input powers of both the probe and the signal pulses in Figs.~4(c) and 4(e) were set to 1 and 0.4 pW, corresponding to 40
and 16 photons, respectively. Circles (squares) represent the experimental data of the probe transmission (phase shift). The blue (red) lines
are the theoretical curves of the probe transmission (phase shift). We note that there was a discrepancy between the experimental data and the
theoretical curves [see Figs.~4(a) and 4(b)]. This deviation is possibly ascribed to the three degenerate Zeeman sublevels of $^{87}{\rm Rb}$
atoms in the experiment, which was not considered in our theoretical model~\cite{GuanZeemanEIT}.

%%%%%%%%%%%%
\FigFour
%%%%%%%%%%%%

In Fig.~4, we also observed a phenomenon of phase jump when the relative phase $\phi_{r}$ was around 4.5 radians. The behavior of the phase jump
closely depends on the initial conditions of the double-$\Lambda$ system. A detailed analysis will be published elsewhere. As the number of the
probe and signal photons decreased, the measurement data became increasingly chaotic, as shown in Figs.~4(c)-4(f). We emphasize that the phase
noise of the beat-note interferometer measured from the fluctuations in the number of detected photons was less than 0.1 radians because each
data point was averaged 4096 times in this measurement~\cite{LoBeatNote}. Therefore, the chaotic data in Figs.~4(c)-4(f) are mainly attributed
to fluctuations in the optical phase and the number of photons of the few-photon pulses, which can be eliminated using a biphoton with a
constant phase difference (i.e., $\phi_{p}-\phi_{s}$ = constant). Although these data possess a large amount of phase noise, the measured values
are still valid considering the considerable phase shift. For instance, in Fig.~4(f), a maximum phase shift of -3.0 radians was observed when
the relative phase $\phi_{r}$ was set to 4.3 radians. In the absence of the signal field, the probe transmission and phase shift were measured
around 22$\%$ and -0.1 radians, respectively. In other words, the signal pulse made the probe pulse acquire a phase shift of order $\pi$ at the
few-photon level has been realized with this double-$\Lambda$ EIT scheme.

%, which are independent of the phases of the applied lasers [see Eq.~(6)],

%%%%%%%%%%%%%%%%%%%%%%%%%%%%%%%%%%%%%%%%%%%%%%%%%%%%%%%%%%%%%%%%%%%%%%%%%%%%%%%%
%%%%%%%%%%%%%%%%%%%%%%%%%%%%%%%%%%%%%%%%%%%%%%%%%%%%%%%%%%%%%%%%%%%%%%%%%%%%%%%%

%\section{CONCLUSION}

In conclusion, we have demonstrated an efficient APM based on a double-$\Lambda$ EIT system at the few-photon level. A nonlinear phase shift of
order $\pi$ induced by a light pulse containing 16 photons was observed in cold $^{87}{\rm Rb}$ atoms with this scheme. The property of the
double-$\Lambda$ medium can be controlled by changing the phases of the applied laser fields. This phase-dependent mechanism has been
investigated in this work. The conventional N-type APM based on EIT at the single-photon level has a theoretical maximum phase shift on the
order of 0.1 radians~\cite{HarrisSLNO,LoFXPM}. Our results shows that this double-$\Lambda$-based APM scheme has the potential to overcome this
limitation, making it feasible to use single photons to induce a conditional phase shift of $\pi$.

%%%%%%%%%%%%%%%%%%%%%%%%%%%%%%%%%%%%%%%%%%%%%%%%%%%%%%%%%%%%%%%%%%%%%%%%%%%%%%%%
%%%%%%%%%%%%%%%%%%%%%%%%%%%%%%%%%%%%%%%%%%%%%%%%%%%%%%%%%%%%%%%%%%%%%%%%%%%%%%%%

%\section{Methods}

%\textbf{Experimental details.}

%%%%%%%%%%%%%%%%%%%%%%%%%%%%%%%%%%%%%%%%%%%%%%%%%%%%%%%%%%%%%%%%%%%%%%%%%%%%%%%%
%%%%%%%%%%%%%%%%%%%%%%%%%%%%%%%%%%%%%%%%%%%%%%%%%%%%%%%%%%%%%%%%%%%%%%%%%%%%%%%%

\section*{ACKNOWLEDGEMENTS}
We acknowledge Bing He and Ray-Kuang Lee for helpful discussions and Jun-Xian Chen for experimental assistance. This work was
supported by the National Science Council of Taiwan under grants no. 99-2628-M-006-011 and no. 100-2628-M-006-001.

\end{document}